# Low-Cost Data Acquisition Card for School-Network Cosmic Ray Detectors

Sten Hansen, Thomas Jordan, Terry Kiper , *Fermi National Laboratory, Batavia IL*

Dan Claes, Gregory Snow, *Univ.of Nebraska, Lincoln, NE*

Hans Berns, T. H. Burnett, Richard Gran, R. Jeffrey Wilkes*, Univ. of Washington, Seattle, WA*

*Abstract*—The Cosmic Ray Observatory Project (CROP) at University of Nebraska/Lincoln and the Washington Area Large-scale Time coincidence Array (WALTA) at University of Washington/Seattle are among several outreach projects siting cosmic-ray detectors at local high schools in cities around North America, to study the origins and interactions of high-energy cosmic rays. In a collaboration between QuarkNet, the outreach program based at Fermilab, CROP, and WALTA, a low-cost data acquisition electronics card has been developed to collect and synchronize the data from each detector site. The cost for each card is under US$500 for parts, functionally replacing much more expensive electronics crates and modules at each high school site. The card has 4 analog discriminator inputs for photo-multiplier tube signals, a 4-channel Time-to-Digital converter for local coincidence and time-over-threshold measurements at 0.75 ns resolution, programmable trigger logic via a CPLD and microcontroller, and a built-in low-cost GPS receiver/antenna module (via external cable) to provide event trigger time stamps at better than 100 ns accuracy. Temperature sensors and a barometer are also integrated to record environmental data along with the counter data. The card connects to any PC or laptop via a standard RS-232 serial port for data output and control. The microcontroller and CPLD are field programmable and therefore make the card functionality flexible and easy to upgrade.

## I. INTRODUCTION

THE QuarkNet [1] DAQ card for school-network cosmic ray detectors [2] provides a low-cost alternative to using standard particle and nuclear physics fast pulse electronics modules. The board, which can be produced at a cost of less than US$500, produces trigger time and pulse edge time data for 2 to 4-fold coincidence levels, via a universal RS-232 serial port interface, usable with any PC.

Individual detector stations, each consisting of 4 scintillation counter modules, front-end electronics, and a GPS receiver, produce a stream of data in the form of ASCII text strings, in a set of distinct formats for different functions.

The card includes a low-cost GPS receiver module, which permits time stamping event triggers to about 50 ns absolute accuracy in UTC, between widely separated sites [3]. The technique used for obtaining precise GPS time was adapted from the procedure developed for the K2K long-baseline neutrino experiment [4].

Development of the QuarkNet DAQ card followed discussions between QuarkNet, CROP and WALTA participants, during and after the SALTA Workshop held as part of the Education and Outreach program at Snowmass-2001 [5]. The need for low-cost but highly capable front-end electronics suitable for very small air shower arrays became clear. A prototype DAQ card previously developed for QuarkNet served as a starting point, but it did not yet include GPS time synchronization or pulse edge time digitization functionality.

CROP [6], WALTA [7] and SALTA [8] all aim to create large-area extensive air shower (EAS) detector arrays for ultra-high energy cosmic rays, by installing mini-arrays of scintillation counter detectors in secondary schools, in the Lincoln, NE, Seattle, WA, and Aspen, CO areas, respectively. Data taken at individual school sites (Fig. 1) are shared via Internet connections and scanned for multi-site coincidence events.

These projects, working together, salvaged plastic scintillators, photomultiplier tubes (PMTs) and associated high voltage supplies from surplus equipment at the CASA detector [10] site at Dugway, UT. Preliminary training of secondary school teachers and students was conducted using NIM crates and fast electronics (discriminators, coincidence and scaler modules) loaned from Fermilab. However, the NIM electronics used did not include GPS timing or a simple computer interface.

The availability of surplus TMC devices purchased for the D0 experiment at Fermilab (Time Measurement Chips [9], pipeline TDCs implemented on an ASIC, and originally developed for ATLAS [11]) made enhanced functionality possible at low cost. Development of the QuarkNet DAQ card represents a significant step in making highly capable fast electronics available at much lower cost than presently available VME and FASTBUS modules, for applications requiring only a few channels.

Manuscript received October 29, 2003. This work was supported in part by the U.S. Department of Energy, Quarknet, and the U.S. National Science Foundation. Contact: wilkes@u.washington.edu.

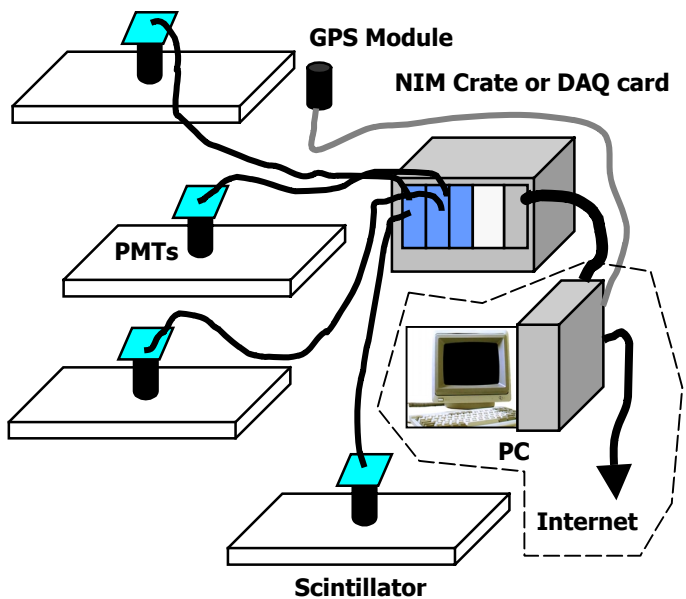

Fig. 1. School-network air shower detector stations. CROP and WALTA use PMTs and counters salvaged from the CASA experiment. Initial installations used NIM crates borrowed from Fermilab PREP, which are being replaced by the DAQ cards described in this paper.

## II. SPECIFICATIONS AND DESIGN

The QuarkNet DAQ card (Fig. 2) was designed to meet the following requirements:
- Four PMT channels with 0-20 dB pre-amplifiers.
- Discriminators with adjustable thresholds.
- Selectable 1- to 4-fold majority trigger logic.
- Leading and trailing edge (relative) arrival times measured with ~1 nanosecond precision.
- Estimate pulse area via time over threshold.
- Onboard scalers for singles and coincidence rates.
- Coincidences stamped with GPS time to ~20 ns.
- Simple digital interface to any PC via serial port.
- Low cost (under US$500 per board for parts).
- Reliable and robust.
- Sufficiently user-friendly for use by high school students.

Nanosecond-scale relative timing on a single board allows rough determination of air shower arrival direction at a given school site via time-of-flight. Inter-site timing on the order of a few tens of nanoseconds, achieved by timestamping triggers with GPS data, is more than adequate for correlation of large-scale air shower events over a many-kilometer scale network of sites.

These specifications were met or exceeded by the design used in Version 2 of the QuarkNet DAQ card. (Version 1, released in 2001, included discriminator and logic functions but not fast relative timing or GPS time synchronization). A functional block diagram of the DAQ card is shown in Fig. 3. Input signals, assumed to be negative-going PMT pulses of height on the order of a few hundred mV and width on the order of a few tens of ns, are first passed through a pre-amplifier stage, with gain field-adjustable by replacing resistors. Cards are produced with gain resistors selected to match the recipient project's PMTs, but this feature permits later adjustment as required.

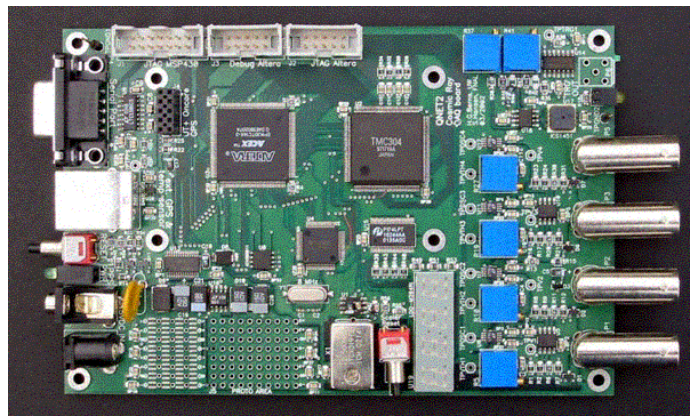

Fig. 2. The QuarkNet DAQ card (Version 2). Scale is indicated by the standard DB-9 connector at upper left.

Following preamplification, signals are sent to discriminators, implemented with voltage comparators and trimpot-adjustable reference voltages.

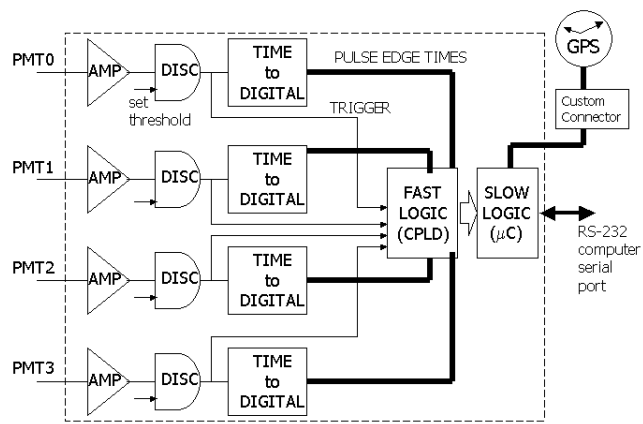

Fig. 3. Hardware block diagram (Version 2).

An Altera EP1K30 CPLD chip implements command-controlled trigger logic, scalers to record singles and coincidence counts, and other fast-logic functions. The user may define the coincidence level, read out or reset counter contents, or perform other functions, by transmitting commands to the board from the host PC. Commands and functions of the CPLD are defined in firmware, which may be altered and updated as needed, by downloading precompiled binary code to the CPLD. Users may, among other functions, set the time window for coincidences from 48 ns to 100 microseconds, and define the effective pipeline length for pulse edge storage in the TDC.

The CPLD is clocked at 41.667 MHz, generated by a 12.50 MHz temperature-compensated quartz oscillator and a 1:3.333 PLL frequency multiplier in series. Thus the basic clock cycle is 24 nsec.

The KEK TMC chip [9] provides a pipeline TDC with 0.75 ns resolution. Leading and trailing edge times from pulses output by the discriminators are stored and made available for readout when a trigger is declared by the CPLD.

The card reports up to one of each (rising and falling) edge per 24 ns clock cycle, for every clock cycle up to the user-specified gate width, thus capturing the full time-structure of air shower events up to as much as 50 microseconds. However, multiple short, closely spaced pulses may be lost if they occur within a single 24 ns clock cycle.

Board-level control functions and host PC interfacing are handled by a Texas Instruments MSP430 microcontroller (MCU) chip. The microcode defining MCU operation can be updated by reprogramming the onboard flash RAM.

In addition to the features mentioned, the card includes an LED display showing the current contents (in hexadecimal) of the coincidence counter, has onboard temperature and barometric pressure sensors, and includes a trigger output connector and manual reset button.

GPS time synchronization data are obtained from a LeadTek GPS receiver module, with receiver electronics integral to the weatherproof antenna module. The GPS receivers used were modified from the manufacturer's stock Model 9532 to provide the 1PPS (1 pulse per second) fast timing signal in addition to the standard ASCII data stream containing time, geographical position and satellite information. A standard DB9 connector was specially adapted to handle the 1PPS signal, add an outdoor temperature sensor, and supply 5VDC power from the DAQ card to the GPS module, eliminating the need for a separate power supply [14].

## III. OPERATION AND PERFORMANCE

From the user standpoint, the DAQ card is very simple to operate. BNC connectors accept coaxial cables from up to 4 PMTs, and the GPS module is connected with its special adapter plug. The user's host computer, which may have a Windows, Mac or Linux OS, is connected via a standard RS232 serial cable. For schools that have relatively new Macintosh models with USB ports only, a commercial USB-to-RS232 adapter can be inserted. The board and GPS unit require a stable 5VDC, 800mA power supply, for example one of the conventional AC adapters available at consumer electronics stores.

The simplest way to operate the board is by opening a terminal window connected to the serial port. Commands can be directly entered via the keyboard, and output viewed (and if desired, captured to a log file) from the screen.

Table I shows the help screen displayed upon board startup, listing the functions implemented. At time of writing (November, 2003), the current CPLD and MCU firmware release is termed Version 2.3.

In typical operation, the user enters commands to define the trigger logic level (1- to 4-fold majority), and enable counting. At the end of a data taking run, counting can be disabled and the trigger count read out. For each trigger, the display shows one or more lines of hexadecimal encoded data, providing time of trigger, pulse leading and trailing edge times relative to the trigger time, and information needed to determine the GPS time of the trigger to 24 ns precision. Other commands can be used to directly interrogate the GPS module, or act upon the counter registers, etc. Users may also prepare a script to implement a sequence of commands, or compile custom software to operate the card directly.

As an example of card performance, Fig. 4 shows a histogram of the measured time separation between pulses from muons penetrating two counters which are in contact (so the actual time of flight is essentially zero). The fitted Gaussian has standard deviation 1.0 ns, only slightly larger than the intrinsic TMC resolution of 0.75 ns.

We have begun development of card interface software [12] using the LabView system from National Instruments [13]. This allows development of user-friendly GUIs by students and teachers, who can make use of extensive resources available from the large LabView user community. In addition to direct data displays, these applications can produce output in many formats, including files formatted for direct use with Excel and other applications familiar to school students and teachers.

While we have focused on cosmic ray air shower detector applications here, the Quarknet DAQ card has additional features, and can be used for experiments which do not require GPS time synchronization, such as muon telescopes, muon decay detectors, and other tabletop particle physics experiments. For example, the card includes the option of setting one channel as a veto in the coincidence logic, as needed in a muon lifetime measurement experiment.

DAQ cards have been extensively tested by the CROP and WALTA groups, and have been used to log useful cosmic ray data from prototype detector sites. Cards have been distributed to over 20 schools and are currently being brought into operation in QuarkNet programs around the USA.

TABLE I
DAQ CARD COMMANDS (HELP SCREEN DISPLAY)

```
Scintillator Card, QNET2  Firmware Ver2.3, 09/09/03    HE=Help
Serial#=1002    uC_Volts=3.3   uC_TempC=26.6   GPS_TempC=235.0    kPa=0

Barometer   - BA=Display, BA bb.b gg.g calibrates kPa Baseline, Gain (See HB).
Counter     - CE=Enable, CD=Disable, Controls TMC Running bit @ CPLD CCR1.
DC          - Display Counters and Control Registers of CPLD, address 0-4.
DF          - Display Scalar Fifo Data (first 12 Bytes as three 32bit numbers).
DG          - Display GPS Date, Time, Position and Status.
DS          - Display Scalar Fifo, Counters 0-3, Triggers, and 1_PPS Time.
DT          - Display Time Control Registers of TMC, address 0-3.
Flash       - FL=Load Binary File, FR=Read SumCheck, FC=Copy_to_CPLD.
GP          - Init Link with GPS unit (GGA=1/sec, RMC=1/sec, disable others).
Help        - HF=Trigger format, HS=Status format, HB=Barometer format.
NA n        - NMEA GPS Data Append (n==1 On),(n!=1 Off), add GPS to output.
NM n        - NMEA GPS Data Echo (n==1 On),(n!=1 Off), (GPS_Baud=9600).
Reset       - RB=TMC+CPLD,  RE=MSP430+TMC+CPLD.
SA n        - SA=Save TMC+CPLD Registers to Flash, (SA 1=Restore Defaults).
SB n        - Set Baud Rate (PC Link), 1=19200, 2=38400, 3=57600, 4=115200.
SN nnnn     - Serial Number(BCD), SN=Display Number, SN nnnn=Store Serial
Number.
ST n        - Send Status Data (n==1 On),(n!=1 Off), (See HF).
TH          - Thermometer Data Display, -40 to 99 degrees C.
WC mm nn    - Write Counter Control Registers CPLD address mm with data nn.
WT mm nn    - Write Time Control Registers TMC address mm with data nn.
```

IV. FUTURE PLANS

We are currently performing intensive testing of the DAQ cards and plan to implement a new firmware version within about 6 months, to repair some firmware bugs and add desired features. The card described is has been distributed to about twenty schools. It is hoped that the present hardware design is sufficiently robust and capable to meet users' needs for some years, with possible future firmware upgrades as applications develop.

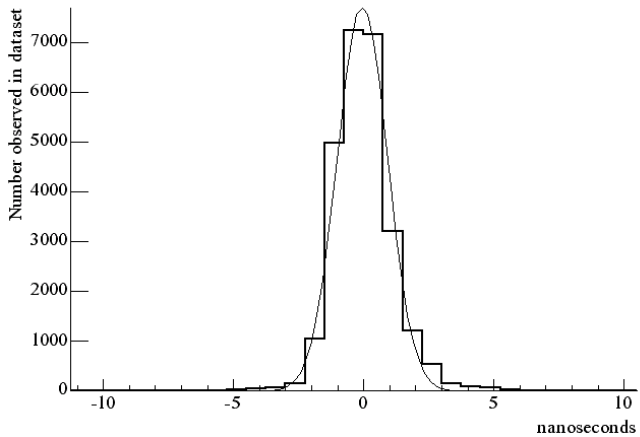

Fig. 4. Histogram of measured time differences between leading edges of pulses from two counters in contact, from a muon telescope experiment at the University of Washington. Actual time differences should be on the order of a few ps. The fitted Gaussian has 1 ns standard deviation, consistent with TMC resolution.

V. ACKNOWLEDGMENT

We wish to thank Mark Buchli, Paul Edmon, Erich Keller, Ben Laughlin, Jeremy Sandler and Jared Kite for valuable assistance in testing and debugging the boards, and the CROP and WALTA teachers for their patience with the development process.